\documentclass[9pt]{article}
\usepackage{spconf}
\usepackage{cite}
\usepackage{graphicx}
\DeclareGraphicsExtensions{.JPG,.eps,.pdf}
\usepackage{amsmath}
\usepackage{amsthm}
\usepackage{amsfonts}
\usepackage{epstopdf}
\usepackage{algpseudocode}
\usepackage{algorithm}
\usepackage{color}
\usepackage{amscd}
\usepackage{bbm}
\usepackage{stfloats}
\usepackage{paralist}
\theoremstyle{remark}

\usepackage{tikz,pgfplots,xcolor}
\usepackage{color}
\usepackage{dsfont}
\usepackage{amssymb}
\newcommand{\be}{\begin{equation}}
\newcommand{\ee}{\end{equation}}
\newcommand{\bea}{\begin{eqnarray}}
\newcommand{\eea}{\end{eqnarray}}
\newcommand{\ben}{\begin{enumerate}}
	\newcommand{\een}{\end{enumerate}}
\newcommand{\beseq}{\begin{subequations}}
	\newcommand{\eeseq}{\end{subequations}}
\algdef{SE}[DOWHILE]{Do}{doWhile}{\algorithmicdo}[1]{\algorithmicwhile\ #1}%

\graphicspath{{./figs/}}
\newlength\figureheight
\newlength\figurewidth
\allowdisplaybreaks

\begin{document}

\title{Deep Learning based Online Power Control for Large Energy Harvesting Networks 
}
\name{Mohit K. Sharma$^1$, Alessio Zappone$^1$,  M\'erouane Debbah$^{1,2}$ and Mohamad Assaad$^1$\thanks{This research has been partly supported by the ERC-PoC 727682 CacheMire project.} }	
\address{$^1$CentraleSupelec, Universit\'e Paris-Saclay, 91192 Gif-sur-Yvette, France\\ $^2$Mathematical and Algorithmic Sciences Lab, Huawei France R\&D, Paris, France}
\author{\IEEEauthorblockN{}\vspace{-0.1in}\\ \\
E-mails: \{mohitkumar.sharma,alessio.zappone\}@l2s.centralesupelec.fr\\
E-mails: merouane.debbah@huawei.com, mohamad.assaad@centralesupelec.fr}

\maketitle
\begin{abstract}
\vspace{-0.05in}

In this paper, we propose a deep learning based approach to design online power control policies for large EH networks, which are often intractable stochastic control problems. In the proposed approach, for a given EH network, the optimal online power control rule is learned by training a deep neural network (DNN), using the solution of offline policy design problem. Under the proposed scheme, in a given time slot, the transmit power is obtained by feeding the current system state to the trained DNN. Our results illustrate that the DNN based online power control scheme outperforms a Markov decision process based policy. In general, the proposed deep learning based approach can be used to find solutions to large intractable stochastic control problems.     
\end{abstract}
\newcounter{MYtempeqncnt}
\vspace{-0.1in}
\section{Introduction}
\vspace{-0.05in}
An energy harvesting (EH) node operates using the energy harvested from the environmental sources\cite{Ku_Comm_Tuts_2016}, e.g., solar, wind, and etc. 
This promises to enable autonomous operation of the next generation wireless networks such as the internet-of-things (IoT)\cite{Centenaro_MWC_Oct2016}. An EH node (EHN) operates under the energy neutrality constraint which requires that at any point in time the total energy consumed by the node, up to that point in time, can not exceed the total amount of energy harvested by the node until that point. Moreover, at a given instant, an EHN with a finite size battery can only store a limited amount of energy. These constraints and the random nature of the EH process make the design of energy management policies a challenging issue in the design of EH systems.  

Depending on the nature of information available about the EH process and wireless channel, the design of energy management policies can be broadly classified in two categories: offline and online methods. In the scenario  when the information about the energy arrivals and the channel state is only causally available\cite{MSharma_TWC_June2018, Baknina_TC_2018}, the energy management policies are termed as online policies. The online policy design problems are stochastic control problems which for an EH network with a large number of nodes suffer from the ``curse-of-dimensionality". In contrast, when the problem horizon is finite and the perfect information about the energy arrivals and the channel state is available in a non-causal fashion, over the entire horizon of the problem, the policies can be designed in an offline manner\cite{kaya_TWC_March2012, Wang_JSAC_Mar2015}. In particular, for a given set of conditions, the offline policy design problem can be posed as a static optimization problem which can be solved efficiently. An offline policy, obtained by solving the static optimization problem corresponding to a given set of realizations of the EH process and the channel, may be suboptimal or even infeasible when used in an online setting. Although designed under different assumptions, both the online and offline policies map a given system state to a feasible output power. Hence, an optimal offline policy also captures the information about the mapping between the system state and the optimal transmit power. Thus, intuitively, by using the offline solution to learn this mapping, it could be possible to design online policies with good performance, for large EH networks.     

In this work, we illustrate the above idea by developing an online power control policy to maximize the time-averaged throughput of a \emph{fading} multiple access channel (MAC) with EH transmitters. In \cite{Wang_JSAC_Mar2015}, the authors designed the throughput optimal offline power control policies for the fading MAC with EH transmitters. On the other hand, the optimal online policies are known only in very restrictive scenarios, such as with binary transmit power levels when the nodes are equipped with infinite size battery \cite{Yang_ISIT_2015} or with unit-sized battery \cite{Blasco_Mar_JSAC2015}\cite{Ianello_CISS_2012}. Furthermore, as the size of the state space grows exponentially with the number of nodes, even obtaining a dynamic programming based numerical solution is not feasible. Hence, the design of optimal online policies for general $K$-user fading MAC where the EHNs are equipped with finite-size battery has remained elusive. 

We design the online policies for fading EH MAC by learning the mapping between the system state and the optimal transmit power using a deep neural network (DNN)\cite{Goodfellow-et-al-2016}. The DNNs, due to its good generalization capabilities, are increasingly being used to improve the performance of wireless communication systems, e.g., for multiple access \cite{Kim_comml_April2018}, power control\cite{Fang_Arxiv_2017} and even for end-to-end reconstruction\cite{Oshea_TGCN_Dec2017} of the data. In our approach, the DNN is trained using the data obtained from offline solutions. The trained DNN is then used to obtain the transmit power vector in an online setting, by feeding the system state as the input. Our results illustrate that the online power control derived using the proposed approach achieves very good performance. The proposed DNN based approach to solve a stochastic control problem with a large state space can be generalized to other domains as well. We note that, in contrast to existing deep learning based approaches \cite{Han_arxiv_2016, Mnih2015_DQN} to solve stochastic control problems our approach differs in the following manner:
\begin{itemize}
	\item The method in \cite{Han_arxiv_2016} is useful only for finite horizon problems and the controller needs to train a DNN for each stage of the problem. On the other hand, our approach is simpler as it needs to train the DNN only once and is applicable to infinite horizon problems.
	\item Compared to \cite{Mnih2015_DQN}, instead of learning the Q-function, the proposed approach directly learns the policy. 
\end{itemize}  
In the next section, we describe the system model which is used for illustrating our approach.
\vspace{-0.1in}	
\section{System Model}
\label{sec:sys}
\vspace{-0.05in}
We consider a time-slotted EH network where $K$ EHNs transmit their data over a fading channel to an AP connected to the mains. The set of transmitters is denoted by $\mathcal{K}\triangleq\{1,2,\ldots,K\}$. Without loss of generality, each slot is assumed to be of unit length and the nodes are synchronized at the slot-level. Each transmission lasts for the entire slot duration\cite{Kapoor_WCL_feb_2017, Liu_TVT_Apr2016}.  In the $n^{\text{th}}$ slot, the complex fading channel gain between the $k^{\text{th}}$ transmitter and the AP is denoted by $g_n^k$. In a given slot, channel between any transmitter and the AP remains constant for the entire slot and changes at the end of the slot. Across the slots, the channel gain between a transmitter and the AP, $g_n^k$, is independent and identically distributed (i.i.d.). 
%

In a slot, the nodes harvest energy according to a general harvesting process with joint probability distribution function denoted by $f_{\mathcal{E}_1,\mathcal{E}_2,\ldots,\mathcal{E}_K }(e_1,e_2,\ldots,e_K)$, where the random variable $\mathcal{E}_k$ denotes the amount of energy harvested by the $k^{\text{th}}$ transmitter and $e_k$ denotes a realization of $\mathcal{E}_k$. The energy harvested by the nodes is independent across the slots. At each node, the harvested energy is stored in a perfectly efficient, finite capacity battery, and at the $k^{\text{th}}$ node the size of the battery is denoted by $B_{\max}^k$. Only \emph{causal} information about the EH process and channel states between the AP and all the nodes is available. In order to illustrate the main idea, we assume that the battery state of all the nodes and the channel states between all the transmitters and the AP is available in a centralized fashion \cite{Wang_JSAC_Mar2015, Arafa_JSAC_Dec2015}. However, as described in Sec.~\ref{sec:SOlution_Method}, the proposed
online policies can be easily adapted for a distributed implementation. 

Let $p_n^k\leq P_{\max}$ denote the amount of energy used by the $k^{\text{th}}$ transmitter in the $n^{\text{th}}$ slot, where $P_{\max}$ denotes the maximum transmit energy which is determined by the RF front end. Further, $\mathcal{P}_n\triangleq\{p_n^k\}_{k=1}^K$
denotes the set of energy levels used, in the $n^{\text{th}}$ slot, by all the transmitters. The battery at the $k^{\text{th}}$ node evolves as  
\be
B_{n+1}^k=\min\{[B_{n}^k+e_n^k-p_n^k]^+, B_{\max}^k\},
\label{eq:battery_evol}
\ee
where $1\leq k\leq K$, and $[x]^+\triangleq\max\{0,x\}$. In the above, $B_{n}^k$ and $e_n^k$ denote the battery level and the energy harvested by the $k^{\text{th}}$ node, at the start of the $n^{\text{th}}$ slot. 
An upper bound on the successful transmission
rate of the EH MAC over $N$ slots is given by\cite{Wang_JSAC_Mar2015}
\be
\mathcal{T}(\mathcal{P}) = \sum_{n=1}^N\log\left(1+\sum_{k\in\mathcal{K}}p_n^kg_n^k\right),
\label{eq:throughput_finite}
\ee
where $\mathcal{P}\triangleq\{\mathcal{P}_n|1\leq n\leq N\}$. Note that, in \eqref{eq:throughput_finite} for simplicity, and without loss of generality, we set the power spectral density of the AWGN at the receiver as unity. 

Our goal in this paper is to find an online energy management policy to maximize the time-averaged throughput. The optimization problem can be expressed as follows
\beseq
\begin{align}
&\max_{\{\mathcal{P}\} } \liminf_{N\to \infty}\frac{1}{N}\mathcal{T}(\mathcal{P}),\\
&\text{s.t. }  0\leq  p_n^k\leq \min\{B_n^k,P_{\max}\}, 
\label{eq:optim_prob_const}
\end{align}
\label{eq:optim_prob}
\eeseq
for all $n$ and $1\leq k\leq K$. The constraint \eqref{eq:optim_prob_const} captures the fact that the maximum energy a node can use in the $n^{\text{th}}$ slot is limited to the minimum of the amount of energy available in the battery, $B_n^k$, and the maximum allowed transmit energy $P_{\max}$. Note that, since the information about the \emph{random} energy arrivals and the channel is only \emph{causally available} and for each node the battery evolves in a Markovian fashion, according to \eqref{eq:battery_evol}, the optimization problem in \eqref{eq:optim_prob} is essentially a stochastic control problem which can be formulated as a Markov decision process (MDP) by discretization of the state space. 
However, for an EH network with large number of nodes or with large battery size at each node, it suffers from the ``curse-of-dimensionality". In the following, we present a deep learning based approach to efficiently solve \eqref{eq:optim_prob}, when only the causal information is available. 
\vspace{-0.1in}
\section{DNN based Online Energy Management}
\label{sec:SOlution_Method}
\vspace{-0.05in}
\begin{figure}[t!]
	\centering
	\includegraphics[width=3in]{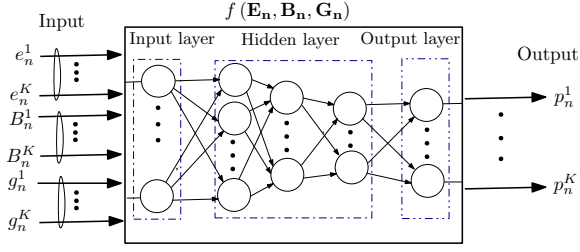}
	\caption{DNN based online energy management policy. In the $n^{\text{th}}$ slot, the DNN maps the system state $(\mathbf{E_n,B_n,G_n})$ to a feasible transmit energy vector.} 
	\label{Fig:DNN_arch}
	\vspace{-0.1in}
\end{figure}
To describe our DNN based approach to solve \eqref{eq:optim_prob} we define some additional notations and  mathematically describe online and offline policy in context of the problem \eqref{eq:optim_prob}.
\vspace{-0.05in}
\subsection{Notations} For the $k^{\text{th}}$ node, let ${\boldsymbol E}_{m:n}^k\triangleq\{e_m^k, e_{m+1}^k,\ldots,e_n^k\}$, ${\boldsymbol B}_{m:n}^k\triangleq\{B_m^k, B_{m+1}^k,\ldots,B_n^k\}$, and ${\boldsymbol G}_{m:n}^k\triangleq\{g_m^k, g_{m+1}^k,\ldots,g_n^k\}$ denote the vectors containing the values of energy harvested, battery state, and the channel state, respectively, in the slots from $m$ to $n$. Further, history up to the start of the slot $n$ is denoted by the tuple $\boldsymbol{H}_n\triangleq\left\{({\boldsymbol E}_{1:n-1}^k, {\boldsymbol B}_{1:n-1}^k, {\boldsymbol G}_{1:n-1}^k)\right\}_{k=1}^K$ and $\boldsymbol{H}_n \in \mathcal{H}_n$, where $\mathcal{H}_n$ is the set of all possible histories up to slot $n$. Also, in the $n^{\text{th}}$ slot, the current state of the system is described by the tuple $\boldsymbol{S}_n\triangleq\{\boldsymbol{E}_n,\boldsymbol{B}_n,\boldsymbol{G}_n\}$, where $\boldsymbol{E_n}\triangleq(e_n^1,e_n^2,\ldots,e_n^K)$, $\boldsymbol{B_n}\triangleq(B_n^1,B_n^2,\ldots,B_n^K)$ and $\boldsymbol{G_n}\triangleq(g_n^1,g_n^2,\ldots,g_n^K)$ are the vectors containing the values of energy harvested, battery state and the channel state, respectively, in the $n^{\text{th}}$ slot, for all the nodes. Further, $\boldsymbol{S}_n\in \mathcal{S}$ where $\mathcal{S}$ denotes the set of all possible states.  
\vspace{-0.05in}
\subsection{Online and Offline Policies}
\vspace{-0.03in}
In the $n^{\text{th}}$ slot, an online decision rule $f_n:\mathcal{H}_n \times \mathcal{S}\to \hat{\mathcal{P}}$ maps the history, $\boldsymbol{H}_n$, and the current state of the system, $\boldsymbol{S}_n$, to a feasible transmit energy vector, $\hat{\mathcal{P}}\in \mathbb{R}_+^K$, which contains the transmit energies for all the nodes. Mathematically, an \emph{online} policy $\mathcal{F}$ is the collection of decision rules, i.e., $\mathcal{F}\triangleq\{f_1,f_2\ldots\}$. 
In contrast, for \emph{offline} policy design problem the horizon $N$ is finite and,  for all the slots, the information about the amount of the energy harvested and the channel state is available non-causally, i.e., $\{E_{1:N}^k\}_{k=1}^K$, $\{G_{1:N}^k\}_{k=1}^K$ are known before the start of the operation. Hence, the stochastic control problem in \eqref{eq:optim_prob} reduces to a static optimization problem whose objective and constraints are deterministic convex functions in the optimization variables $p_k^n$. Hence, in the offline scenario, \eqref{eq:optim_prob} reduces to a convex optimization problem which can be solved efficiently using the iterative algorithm presented in \cite{Wang_JSAC_Mar2015}, with complexity equal to $\mathcal{O}\left(KN^2\right)$. 
\vspace{-0.05in}
\subsection{DNN based Online Policy} 
\vspace{-0.03in}
To develop the DNN based online energy management policy, we first note that, due to finite size of the state and action spaces of the problem, the optimal policy for the problem \eqref{eq:optim_prob} is a Markov deterministic policy\cite[Thm. 8.4.7]{Puterman_MDP_2014}, i.e., $\mathcal{F}\triangleq\{f,f\ldots\}$ where $f:\mathcal{S}\to \hat{\mathcal{P}}$. Hence, the optimal online energy management policy can be obtained by finding a decision rule which maps the current state of the system $\boldsymbol{S_n}$ to an optimal transmit energy vector for problem \eqref{eq:optim_prob}. Furthermore, for a finite horizon problem, an offline policy also represents a mapping from the current state to a feasible transmit energy vector, i.e., the optimal offline policy maps a $(\boldsymbol{E},\boldsymbol{B}, \boldsymbol{G})$ tuple to $\hat{\mathcal{P}^*}$. Since a DNN is a universal function approximator\cite{Goodfellow-et-al-2016}, provided it contains a sufficient number of neurons, we propose to use a DNN to learn the optimal decision rule, by using the solution of the offline policy design problem to train a DNN. In a given slot, using the proposed online scheme, the optimal transmit energy vector can be obtained by feeding the current state of the system as the input to the trained DNN. Our approach is illustrated in Fig.~\ref{Fig:DNN_arch}. Next, we briefly describe the architecture of the DNN and the procedure used for training the DNN.
\vspace{-0.05in}
\subsection{DNN Architecture and Training}  
\vspace{-0.03in}
We adopt a feedforward neural network\cite{Goodfellow-et-al-2016} with fully-connected layers, whose input layer contains $3K$ neurons, one corresponding to each input. A $3K$-length vector, containing the system state, is fed to DNN as input which is then processed by $h+1$ layers ($h$ hidden layers and an output layer) to output a $K$-length vector of transmit energies. The number of processing units, usually termed as neurons, at the $j^{\text{th}}$ layer, where $1\leq j\leq h+2$, is denoted by $N_j$. Note that, $N_1=3K$ and $N_{h+2}=K$. The output of the $n^{\text{th}}$ neuron of the $j^{\text{th}}$ layer is computed as 
$I_j(n) =  F_{j,n}\left(\boldsymbol{W}_{j,n}^T\boldsymbol{I}_{j-1}+b_{j,n}\right)$,  
where $\boldsymbol{I}_{j-1}$ denotes the output of the $(j-1)^{\text{th}}$ layer, which is fed as input to the $j^{\text{th}}$ layer. Also, $\boldsymbol{W}_{j,n}\in \mathbb{R}^{N_{j-1}}$, $b_{j,n}\in \mathbb{R}$, and $F_{j,n}$ denote the weights, bias, and the nonlinear activation function, respectively, for the $n^{\text{th}}$ neuron of the $j^{\text{th}}$ layer.  

The DNN can learn the optimal mapping between the system state and the transmit energy vector, by appropriately adjusting the weights and biases of the neurons in the network. The weights $\boldsymbol{W} = \{\{\boldsymbol{W}_{j,n}\}_{n=1}^{N_j}\}_{j=1}^{h+2}$and biases $\boldsymbol{b}=\{\{b_{j,n}\}_{n=1}^{N_j}\}_{j=1}^{h+2}$ of the neurons of a DNN can be tuned by minimizing a loss function over a training set which is a set of data points for which the optimal mapping is already known. In particular, the training process minimizes the average loss, over the entire training set, which is defined as
\be
L_{\text{av}}(\boldsymbol{W}, \boldsymbol{b})=\frac{1}{N_{\text{data}}}\sum_{\ell=1}^{N_{\text{data}}}L\left(\hat{\mathcal{P}^*},\boldsymbol{I}_{h+2}(\boldsymbol{W},\boldsymbol{b})\right),
\label{eq:loss_function}
\ee   
where $L(\cdot)$ denotes a loss function which is a metric of distance between the desired output and the output of the DNN, and $N_{\text{data}}$ denotes the number of data points in the training set. The training proceeds by iteratively minimizing the loss in  \eqref{eq:loss_function}, using the gradient based methods over the training data set. The details related to the loss function, training method, and the batch size used in this work are presented in Sec.~\ref{sec:sim}.  

Note that, the training data is generated by solving \emph{multiple instantiations} of the offline problem, each corresponding to a different realization of $\{\boldsymbol{E}_{1:N}^k, \boldsymbol{G}_{1:N}^k\}_{k=1}^K$. The training data contains the tuples of the form $\{\left(\boldsymbol{E},\boldsymbol{B},\boldsymbol{G}\right),\boldsymbol{P}\}$, where $\left(\boldsymbol{E},\boldsymbol{B},\boldsymbol{G}\right)$ and $\boldsymbol{P}$ represent the input to the DNN and the desired output, respectively. We note that, once the DNN is trained, the proposed policy entails a low computational cost to compute the transmit powers, i.e., it requires $\sum_{j=1}^{h+2}N_jN_{j-1}$ multiplication. Furthermore, the proposed policy can be easily implemented in a distributed fashion by broadcasting the result of centralized training process (optimal weights and biases) to all the nodes.
 
\vspace{-0.1in}
\section{Simulation Results}
\label{sec:sim}
\vspace{-0.05in}
We consider an EH MAC with $K=5$ EH transmitters. The channel between each EHN and the AP is i.i.d. Rayleigh faded with average channel gain equal to unity. Each EHN harvests energy according to a nonnegative truncated Gaussian distribution with mean $m$ and variance $v$, independently of the other nodes. The size of the battery at each transmitter is $B_{\max}=20$, and the maximum amount of energy allowed to be used for transmission in a slot is $P_{\max}=15$. Note that, the unit of energy is $10^{-2}$ J. The metric used to analyze the performance is the total rate obtained per slot (RPS).

We use a DNN with an input and output layer containing $3K$ and $K$ neurons, respectively. It consists of $30$ hidden layers, with first hidden layer containing $30K$ neurons. Each subsequent odd indexed hidden layer contains the same number of neurons as the previous even indexed layer, i.e., $N_j=N_{j-1}$ for $j\in\{3,\ldots,31\}$. For each even indexed hidden layer the number of neurons is decreased by $2K$, i.e., $N_{j}=N_{j-1}-2K$ for $j\in \{4,\ldots,30\}$. We note that, the input layer has the index $1$, and the first hidden layer and the output layer having index $2$ and $32$, respectively. The activation function used is Leaky rectified linear unit (ReLu). To train the network we use the mean-square error as the loss function. Training data is generated by solving $10^4$ instantiations of the offline problem with the horizon length $N=20$. Thus, the training dataset contains $2 \times 10^5$ datapoints, out of which $40000$ datapoints are used for validation. The performance is evaluated by computing the RPS over $10^6$ slots. For these $10^6$ slots, instantiations of the EH process and the channel are generated independently of the training data.    
\begin{table}[t!]
	\renewcommand{\arraystretch}{1.3}
	\caption{Performance of the DNN based policy for an EH MAC with $K=5$ users and $v=3.5$. Performance of the offline policy corresponds to 100\%. }   
	\label{table_I}
	\centering
	\begin{tabular}{|c|c|c|c|}
		\hline
		 \begin{tabular}[x]{@{}c@{}}Mean\\ (m)\end{tabular} &  \begin{tabular}[x]{@{}c@{}}Offline Policy\\ (RPS in nats)\end{tabular} & \begin{tabular}[x]{@{}c@{}}DNN policy\\ (RPS in nats)\end{tabular} & \begin{tabular}[x]{@{}c@{}}DNN policy\\ (Percentage )\end{tabular}\\
		\hline
		$4$  &  3.4907 &  3.1498 &  90.23\%\\ 
		\hline
		$5$  &  3.6564 &  3.3107 & 90.54\%\\
		\hline
		$6$  &   3.7877 & 3.4410 & 90.84\% \\
		\hline
		$7$  &  3.8922 & 3.5102 & 90.18\%  \\
		\hline
		$8$  &  3.9740 & 3.6146 & 90.95\% \\
		\hline
		$9$  &  4.0407 & 3.5676& 88.29\% \\
		\hline
		\end{tabular}
\end{table}

Table~\ref{table_I} shows the performance of the proposed DNN based policy. The last column of the table presents the RPS as the percentage of the throughput achieved by the offline policy. It can be observed that the proposed policy achieves roughly $90\%$ of the throughput obtained by the offline policy. We note that, since an offline policy is designed using non-causal information, the proposed policy can not  achieve the throughput obtained by the optimal offline policy. Note that, the MDP formulation of this problem is computationally intractable, due to \emph{state space of the size} of order $10^{12}$, even with the channel gains quantized to just $8$ levels. Due to lack of space, we omit the comparison of the proposed approach against the deep Q-learning based policies. It will be presented in the longer version of the paper.  

Results in Table~\ref{table_P2P_mean10} compares the performance of the proposed DNN based policy against MDP, for point-to-point links, i.e., $K=1$, with mean $m=10$. The proposed DNN based policy achieves approximately $98$~\% of the time-averaged throughput achieved by the offline policy. It is interesting to note that the proposed policies outperform the online policies designed using the MDP which achieves only approximately $84$~\% of the throughput achieved by the offline policy. Theoretically, an online policy designed using MDP achieves the optimal performance. However, the performance of MDP policy degrades due to quantization of the state and action spaces. On the other hand, the proposed DNN based policy operates with continuous state and action spaces. 
\begin{table}[t!]
	\renewcommand{\arraystretch}{1.3}
	\caption{Performance of the DNN based online policy for a point-to-point link with $m=10$. }   
	\label{table_P2P_mean10}
	\centering
	\begin{tabular}{|c|c|c|c|}
		\hline
		\begin{tabular}[x]{@{}c@{}}Variance\\ (v)\end{tabular} &  \begin{tabular}[x]{@{}c@{}}Offline Policy\\  (RPS in nats)\end{tabular} & \begin{tabular}[x]{@{}c@{}}DNN Policy\\ (Percentage )\end{tabular} & \begin{tabular}[x]{@{}c@{}}MDP Policy\\ (Percentage )\end{tabular}\\
		\hline
		$1$  &  2.0434 & 98.41\%  & 83.32\% \\ 
		\hline
		$2$  &  2.0375 & 98.56\%  & 83.60\% \\
		\hline
		$3$  &   2.0372 & 98.38\% & 83.32\% \\
		\hline
		$4$  &  2.0347 & 95.85\% & 83.37\% \\
		\hline
		$5$  &  2.0310 & 97.72\% & 83.29\% \\
		\hline
		$6$  &  2.0284 & 98.22\%  & 83.21\% \\
		\hline
	\end{tabular}
\end{table}
\vspace{-0.1in}
\section{Conclusions}
\label{sec:conclude}
\vspace{-0.05in}
In this paper, we proposed a noble deep learning based method to solve the stochastic control problems, using the solution of the offline problem for training the DNN. The offline problems are static optimization problems which, compared to the original stochastic control problem, often can be solved efficiently. We illustrated our approach by developing an online power control policy to maximize the throughput of an EH based fading MAC. We trained the DNN using the training data generated by solving the offline power control problem. It is observed through simulations that the proposed approach provides a very good performance. 

\bibliographystyle{IEEEtran}
{\bibliography{bibs/IEEEabrv,bibs/bibJournalList,bibs/references}}

\end{document}